# Single-pulse and two-pulse echoes at multipulse excitation mode in multidomain magnetic materials

**A.Akhalkatsi [a], T.Gegechkori [b], G.Katalandze[a], G.Mamniashvili [b,*], Z.Shermadini [a]**

[a] I.Javakhishvili Tbilisi State University
[b] E.Andronikashvili Institute of Physics, Georgian Academy of Sciences

[*] Corresponding author.
E-mail address: gmamnishvili@yahoo.com
(G.I.Mamniashvili)

## Abstract

Properties of single-pulse and two-pulse echoes and their secondary echo signals as well as the role of reversible relaxation in the observed decays of echo signals at the multipulse excitation mode in a number of multidomain magnetic materials were studied.

The single-pulse echo (SPE) is a resonance response of spin-system on the application of solitary exciting radiofrequency (RF) pulse arising at a time approximately equal to the pulse duration after its termination. Though SPE was discovered by Bloom yet in 1955 soon after Hahn's discovery of the famous two-pulse echo (TPE) phenomena, the mechanism of its formation appeared to be far more complicated, as compared with the TPE one, and continues to attract the attention of researchers [1].

SPE formation mechanisms could be conditionally classified on co-called edge-type ones when RF pulse edges act like two RF pulses in the Hahn method – such as the nonresonant mechanism [2,3] and the distortion mechanism [4], and also internal mechanisms due to the presence in the spin-system dynamics particular nonlinearities, as example, connected with a strong dynamic frequency shift of the NMR frequency or with a nonlinear dynamics of nuclear spins due to the simultaneous presence of large Larmor and Rabi inhomogeneous broadenings of the NMR line [1].

In this work we consider in more details the so-called multipulse mechanism of SPE formation, presented in work [1], for systems with both types of frequency inhomogeneities of NMR lines. An important example of such systems is multidomain magnets, for example, lithium ferrite. Earlier in work [5] it was investigated by us the peculiarities of SPE formation in this magnetics. It was established that its properties sharply differ from SPE properties in cobalt where it was formed by the distortion mechanism. Therefore, the conclusion was made on the possible effectiveness of new internal mechanism of the SPE formation in lithium ferrite. But in work [5] the SPE formation mechanism in lithium ferrite was not finally established what was made in works [6,7] where it was shown that SPE and its secondary signals properties in lithium ferrite were well described by the multipulse mechanism of formation [1].

Shakhmuratova et al. [1] used the formalism of statistic tensors to perform a theoretical investigation of the formation of SPE and its secondary signals in the presence of large Larmor and Rabi inhomogeneous broadenings of the NMR line, which, e.g. takes place in multidomain ferromagnets, when the repetition period of RF pulse T satisfies the following inequality for the characteristic relaxation parameters:

$$T_3 \ll T_2 < T < T_1. \tag{1}$$

Where $T_1$ is the spin-lattice relaxation time, $T_2$ is the transverse irreversible relaxation time, $T_3$ characterizes the transverse reversible relaxation time ($T_3 \sim 1/\sigma$, where $\sigma$ is the halfwidth at halfmaximum of the inhomogeneously broadened line).

Under these conditions, the RF cycles are applied to a nonequilibrium spin system, and in the end of each period T we should take into account only the longitudinal component of the nuclear magnetization as the initial condition for the consideration of the dynamics of the spin system

It was shown that the dephasing of the spin system, which is accumulated in the course of n-fold repetition of the pulsed excitation, is recovered during a time interval following the (n+1) th "read-out" pulse in the multipulse sequence, which leads to the formation of an SPE and its secondary signals at time moments multiple of the duration of the RF pulse τ after the termination of the "read-out" pulse.

In work [7] it was shown that obtained in [1] by the statistical tensors method expressions for the transverse components of nuclear magnetizations could be also obtained in the framework of the usual classical-approach by solving the system of Bloch equations from [2], where it was allowed for both types of inhomogeneous broadenings of NMR lines and condition (1). But the approach of work [8] using the Mims transformation matrix method [9], appeared to be more visual.

Let us consider the case when a local static magnetic field $\mathbf{H}_n$ is directed along $\mathbf{Z}$ axis and RF field is along $\mathbf{X}$ axis of the rotating coordinate system (RCS) when an effective magnetic field modulus in RCS is given by the expression:

$$H_{eff} = \frac{1}{\gamma_n}\sqrt{\Delta\omega_j{}^2 + \omega_1{}^2} \text{, where } \psi_j \text{ is angle between } \mathbf{H}_{effj} = \frac{1}{\gamma_n}(\Delta\omega_{nj}\hat{\mathbf{z}} + \omega_1\hat{\mathbf{y}}) \tag{2}$$

($\hat{\mathbf{z}}$, $\hat{\mathbf{y}}$ are unit vectors in RCS) and $\mathbf{Z}$ axis is defined by the relation:

$$\sin\psi_j = \frac{\omega_1}{\Delta\omega'_j} \text{ ; } \quad \cos\psi_j = \frac{\Delta\omega_j}{\Delta\omega'_j}.$$

Where $\Delta\omega'_j = \sqrt{\Delta\omega_j{}^2 + \omega_1{}^2}$ is the angular velocity of the precession of the j-th isochromate around $\mathbf{H}_{eff}$, $\Delta\omega_j = \omega_{NMR} - \omega_{RF}$ is the detuning for the j-th isochromate, $\omega_1 = \gamma_n\eta H_1$ is the pulse amplitude in frequency units, $\eta$ is the gain of the RF field, $\gamma_n$ is the nuclear gyromagnetic ratio.

Time t characterizes time interval after a pulse termination and is counted from the back front of RF pulse.

The transformation matrix (R) describing the rotation of the magnetization vector $\mathbf{m} = (m_x; m_y; m_z)$ around $\mathbf{H}_{eff}$ is [9]:

$$(R) = \begin{pmatrix} S_\psi{}^2 + C_\psi{}^2 C_\theta & -C_\psi S_\theta & S_\psi C_\psi(1 - C_\theta) \\ C_\psi S_\theta & C_\theta & -S_\psi S_\theta \\ S_\psi C_\psi(1 - C_\theta) & S_\psi S_\theta & C_\psi{}^2 + S_\psi{}^2 C_\theta \end{pmatrix}.$$

Where $C_\psi$, $S_\psi$, $C_\theta$ and $S_\theta$ stand for $\cos\psi$, $\sin\psi$, $\cos\theta$, $\sin\theta$ and $\Psi = \mathrm{tg}^{-1}\left(\dfrac{\omega_1}{\Delta\omega_j}\right)$ is an angle between the effective field $\mathbf{H}_{\text{eff}}$ and $\mathbf{Z}$ axis, $\theta$ is the angle by which the magnetization turns about $\mathbf{H}_{\text{eff}}$ during the pulse time $\tau$: $\theta = \gamma \cdot \mathbf{H}_{\text{eff}} \cdot \tau$, where $\mathbf{H}_{\text{eff}}$ is given by (2).

Let us consider firstly the case of single-pulse excitation. Let

$$X_j = m_{xj}/m \; ; Y_j = m_{yj}/m \; ; Z_j = m_{zj}/m \quad \text{and} \quad \boldsymbol{\mu} = (X_j; Y_j; Z_j),$$

where m is the magnetization modulus. If before the excitation by a RF pulse the nuclear spin system was at equilibrium condition, $\boldsymbol{\mu}_{eq} = (0;0;1)$ then the result of RF pulse action is presented by the $\boldsymbol{\mu} = (R) \cdot \boldsymbol{\mu}_{\text{eq}}$.

After the termination of RF pulse isochromates precess freely around the $\mathbf{Z}$ axis what is described by the matrix:

$$(R_\varphi) = \begin{pmatrix} C_\varphi & -S_\varphi & 0 \\ S_\varphi & C_\varphi & 0 \\ 0 & 0 & 1 \end{pmatrix},$$

where $\varphi = \Delta\omega_j t$ is the angle of rotation of an isochromate around the $\mathbf{Z}$ axis, and t is the time elapsing from the trailing edge of pulse. Therefore, we have finally:

$$\boldsymbol{\mu}_1 = \left(R_\varphi\right)\left(R\right)\boldsymbol{\mu}_{\text{eq}} \begin{pmatrix} C_\varphi S_\psi C_\psi(1-C_\theta) + S_\varphi S_\psi S_\theta \\ S_\varphi S_\psi C_\psi(1-C_\theta) - C_\varphi S_\psi S_\theta \\ C_\psi{}^2 + S_\psi{}^2 C_\theta \end{pmatrix}$$

Expressions for the magnetization compounds in $\boldsymbol{\mu}_1$ coincide with ones obtained in [1] and [2] for the case of single-pulse excitation.

Let us find now the effect of n-fold repetition of the pulsed excitation in the frameworks of model [1] when before the next RF pulse of a train only the longitudinal component of nuclear magnetization remains. It is not difficult to prove by successive matrix multiplication that the expression for equilibrium nuclear magnetization $\boldsymbol{\mu}_{\text{eq}}$ before the final "read-out" (n +1)th pulse is:

$$\boldsymbol{\mu}_n = \left(C_\psi{}^2 + S_\psi{}^2 C_\theta\right)^n \boldsymbol{\mu}_{eq}.$$

Then the result of excitation by the "read-out" pulse and following free precession of magnetization is described directly as for the initial conditions:

$$\boldsymbol{\mu}_{n+1} = \left(R_\varphi\right)\left(R\right)\boldsymbol{\mu}_n = \left(C_\psi{}^2 + S_\psi{}^2 C_\theta\right)^n \begin{pmatrix} C_\varphi S_\psi C_\psi(1-C_\theta) + S_\varphi S_\psi S_\theta \\ S_\varphi S_\psi C_\psi(1-C_\theta) - C_\varphi S_\psi S_\theta \\ C_\psi{}^2 + S_\psi{}^2 C_\theta \end{pmatrix}.$$

These expressions coincide with the ones obtained in [1] using the formalism of statistical tensors. The nth degree multiple has a simple physical meaning of a longitudinal nuclear magnetization created by the n elementary pulses of a multipulse train reflecting the spin system memory to the excitation. The expressions for the SPE and its secondary echo signals using similar expressions for nuclear magnetization vectors were obtained in [1].

It is very easy to carry out this approach also for the case of periodic two-pulse excitation which is of interest for describing the secondary echo signals of two-pulse echo (TPE) in the investigated systems.

It is easy to see that the expression for $\boldsymbol{\mu}_{n+1}$ in the case of single-pulse periodic excitation could be presented as:

$$\boldsymbol{\mu}_{n+1} = Z^n \cdot \begin{pmatrix} C_\varphi X - S_\varphi Y \\ S_\varphi X + C_\varphi Y \\ Z \end{pmatrix}, \qquad (3)$$

where X,Y,Z are the components of nuclear magnetization immediately after the single-pulse excitation. Similarly to this for the periodic two-pulse train allowing for condition (1) one could easily obtain components of nuclear magnetizations after the (n+1)th "read-out" pair of RF pulses in the form of (3) where this time X,Y,Z are components of nuclear magnetization immediately after the solitary two-pulse excitation which could be readily obtained from relation: $\boldsymbol{\mu}_1 = R \cdot R_\varphi \cdot R \cdot \boldsymbol{\mu}_{eq}$ and

$$X = (S_\psi^{\,2} + C_\psi^{\,2} S_\theta) \cdot [C_\varphi S_\psi C_\psi (1-C_\theta) + S_\varphi S_\psi S_\theta] - C_\psi S_\theta \cdot (S_\varphi S_\psi C_\psi (1-C_\theta) - C_\varphi S_\psi S_\theta] +$$
$$\qquad S_\psi C_\psi (1-C_\theta) \cdot (C_\psi^{\,2} + S_\psi^{\,2} C_\theta)$$

$$Y = C_\psi S_\theta \cdot [C_\varphi S_\psi C_\psi (1-C_\theta) + S_\varphi S_\psi S_\theta] + C_\theta \cdot [S_\varphi S_\psi C_\psi (1-C_\theta) - C_\varphi S_\psi S_\theta] -$$
$$\qquad S_\psi S_\theta \cdot (C_\psi^{\,2} + S_\psi^{\,2} C_\theta)$$

$$Z = S_\psi C_\psi (1-C_\theta) \cdot [C_\varphi S_\psi C_\psi (1-C_\theta) + S_\varphi S_\psi S_\theta] + S_\psi S_\theta \cdot [S_\varphi \cdot S_\psi C_\psi (1-C_\theta) - C_\varphi S_\psi S_\theta] +$$
$$\qquad [C_\psi^{\,2} + S_\psi^{\,2} C_\theta]^2$$

Here, as in the case of single-pulse excitation the presence of nth degree multipler reflects the memory of spin system on its excitation by the train of double RF pulses what results similarly to [1] in the formation of secondary signals of TPE.

Let us present the main experimental results obtained in this work. Measurements were carried out by the NMR spectrometer and using diamagnetically diluted lithium ferrite $Li_{0.5}Fe_{2.35}Zn_{0.15}O_4$, Co and $Co_2MnSi$ described in work [5].

Let us firstly present results of comparative analysis of SPE and TPE amplitudes and their secondary signals in the function of repetition period T of exciting RF pulses.

In Fig.1 it is presented the oscillograms of SPE and its secondary signals at the optimal value of repetition period of RF pulse, and Fig.2 indicate the oscillograms of TPE and its secondary signals at the optimal value of repetition period T of RF pulses and Fig.3 and 4 show dependences of their intensities on T in lithium ferrite.

The particular attention was paid to the investigations of SPE at large T when the condition of single-pulse excitation $T \gg T_1$ is obeyed what was made using storage oscilloscopes.

Oscillograms in Fig.5 show the process of complete disappearance of SPE signal in lithium ferrite in the limit of single-pulse excitation and the absence any contribution in the SPE intensity due to the distortion mechanism. From all studied by us magnetics only in lithium ferrite it was not find such contribution. As it is known [4], the intensive SPE signals at single-pulse excitation observed in ferrometallic hexagonal cobalt and in intermetallic $Co_2MnSb$. A similar sufficiently intensive SPE signal was observed in halfmetallic NiMnSb and a much weaker signal in such "bad" metals as manganites. For example, let us present the SPE and TPE dependences for Co and $Co_2MnSi$ on T in Fig.6 and 7, correspondingly. In these materials the main formation SPE mechanism is the distortion mechanism [4]. The effectivity of this mechanism in these materials was confirmed by us by the effect of SPE enhancement at pulsed frequency distortion of the exciting RF pulse trailing edge introduced similarly to work [10]. Similar excitation in the case of lithium ferrite resulted only in the suppression of SPE signal in this material.

From the obtained results it could be noted the correlation of the physical properties of studied by us magnets with intensities of observed SPE signals in them, formed by the distortion mechanism, beginning from the most intensive signal in ferrometall Co down to a weak in "bad" ferrometal manganite and its full disappearance in magnetic dielectric lithium ferrite. This problem could present some interest from the point of view of the practical applications of SPE for the aims of magnetic materials science.

The second interesting moment is the full disappearance of the SPE signal in lithium ferrite in the limit of single-pulse excitation. This result is in an apparent disagreement with the conclusion drawn in work [11] where an echo-like signal observed in glycerin at resonance excitation and with a perfect RF pulse was interpreted as a SPE signal formed by nonresonance mechanism [2] which is clearly described in the frames of the SPE vector model [3]. It is interesting to note that a similar dependence for the echo-like response in the aqueous solution of $MnCl_2$ was also obtained in [12].

A possible explanation of the apparent discrepancy is a low sensitivity of our incoherent spin-echo spectrometer [5] unsufficient to record the SPE signal in lithium ferrite at the single-pulse excitation limit. In works [10, 11] corresponding signals were obtained using coherent spin-echo NMR spectrometers with signal averaging. Another important explanation could be that signals observed in works [10,11] are accordingly [13] a part of an oscillatory free-induction decay (OFID) and effectively averaged down to zero in the case of systems with a large inhomogeneous broadening of NMR lines, such case as in lithium ferrite, what was shown, for example, in [1] (Fig.1 in [1]). This question is also to be studied additionally.

The other interesting problem is the investigation of influence of reversible and irreversible transverse relaxations on the SPE amplitude.

In work [2] it was shown that with a help of SPE signal one could define the transverse relaxation time $T_2$ in magnetics of $MnFe_2O_4$ type. The transverse relaxation time $T_2$ measured by this method does not coincide with the value measured by the TPE method [2]. As it was turned out the SPE signal dependence on the duration $\tau$ of the exciting RF pulse is defined not only by the irreversible relaxation but also by other mechanisms related with the SPE formation, for instance, by the inhomogeneous broadening of the NMR line [14].

Lithium ferrite is the interesting object to study the relaxation processes at the SPE formation because of the absence of the contribution of distortion mechanism in it, and the SPE formation process is only described by the multipulse mechanism the physical nature of which was above discussed.

In Fig.8, 9 it is presented the experimental dependences of SPE main $I_1^s$ and secondary $I_2^s$ signals and dependences of the TPE main and secondary signals ($I_1$, $I_2$), correspondingly. As it is known, in the case of the effectivity of the distortion mechanism the SPE and TPE signals are change similarly at the variation of their excitation conditions [3] and in this case usually holds following relation between transverse relaxation times of TPE ($T_2$) and SPE ($^s T_2$):

$$^s T_2 = (0.5 \div 0.8)\, T_2$$

what is not the case for lithium ferrite where $^s T_2$ is about 30 times shorter as compared with $T_2$, as example, a $f_{NMR}$=73.4 MHz, $T_2$=1200 µs and $^s T_2$=40 µs .

So, the results of experimental investigations of SPE transverse relaxation rate in lithium ferrite speak in the favor of the essential influence of reversible transverse relaxation on the SPE relaxation rate in lithium ferrite in correspondence with works [2, 14].

In conclusion, let us note that in the present work it was studied the SPE properties in the multipulse excitation mode in a number of magnetics. It was discussed the role of different SPE formation mechanisms in connection with the interpretation of the obtained experimental results. It was studied the decay of SPE and TPE intensities on the duration of RF pulses and the time interval between them, correspondingly. It was discussed the role of reversal transverse relaxation for the interpretation of the observed unusually short transverse relaxation times of the SPE signals in lithium ferrite.

This work was supported by the STCU Grant Ge-051 (J).

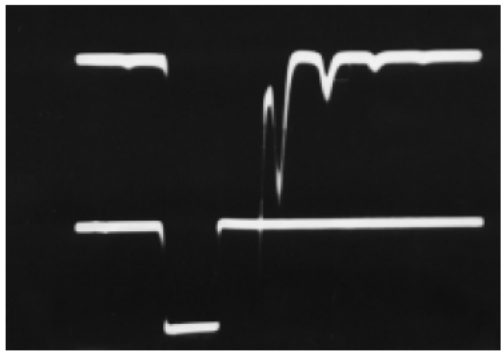

Fig.1

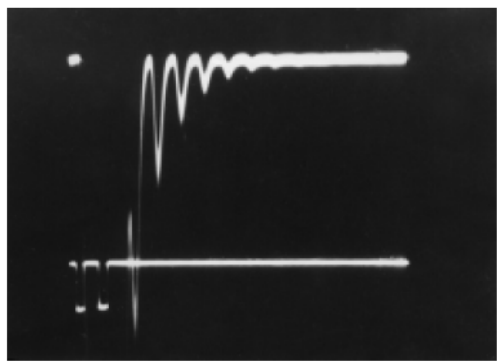

Fig.2

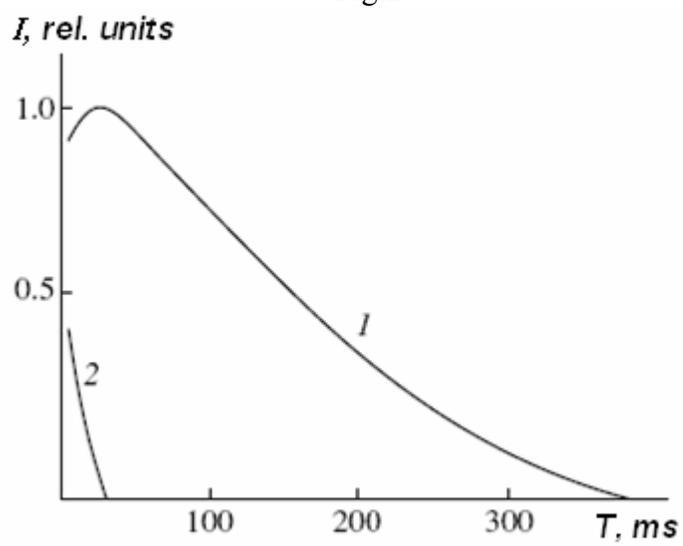

Fig.3

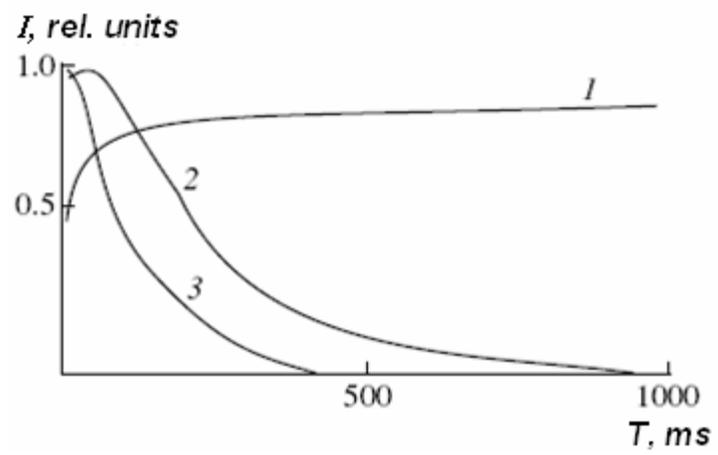

Fig.4

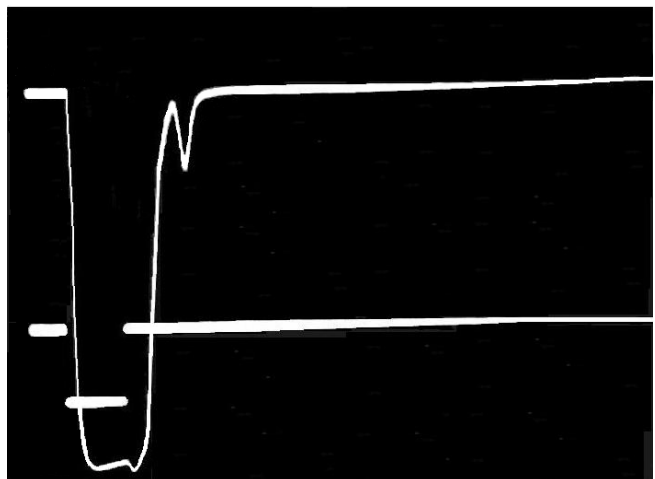

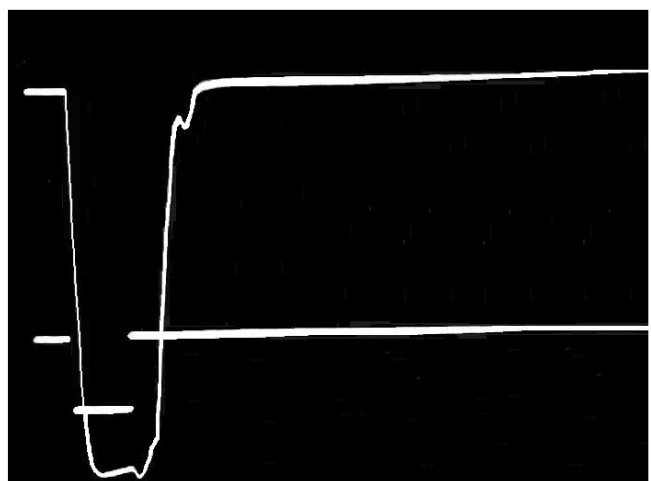

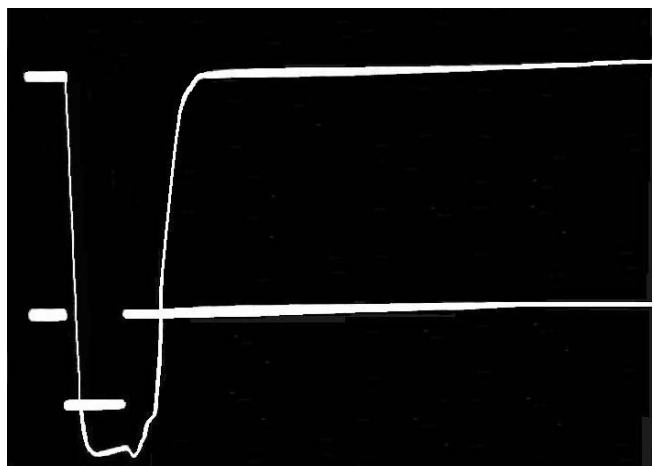

Fig.5

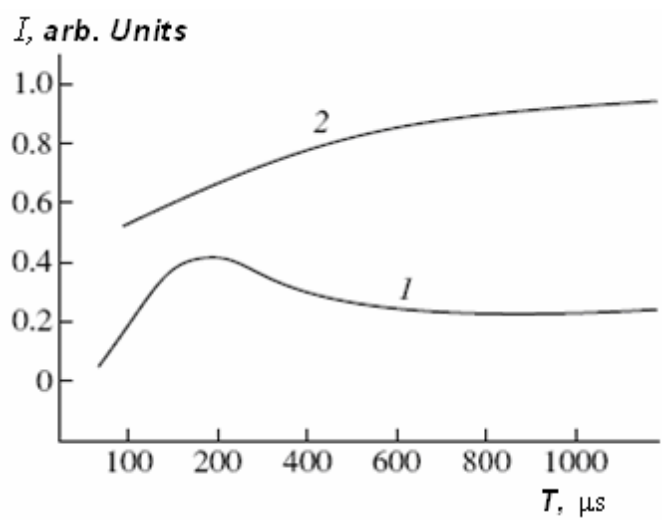

Fig.6.

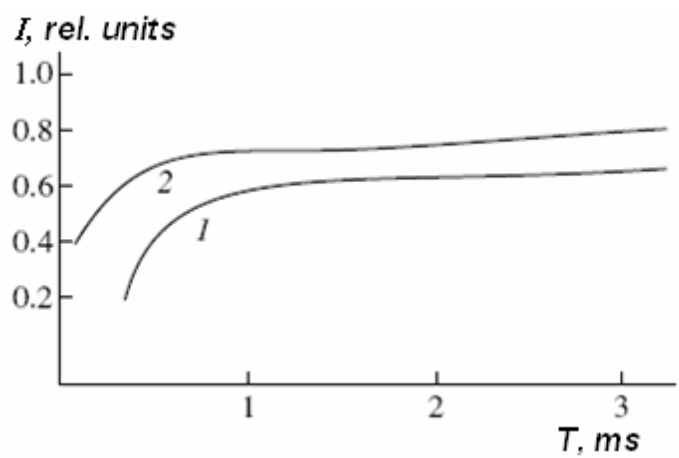

Fig.7.

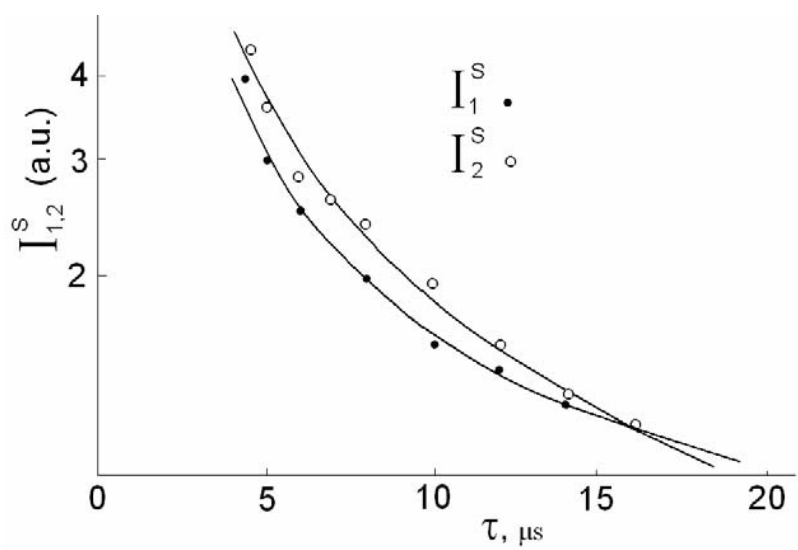

Fig.8.

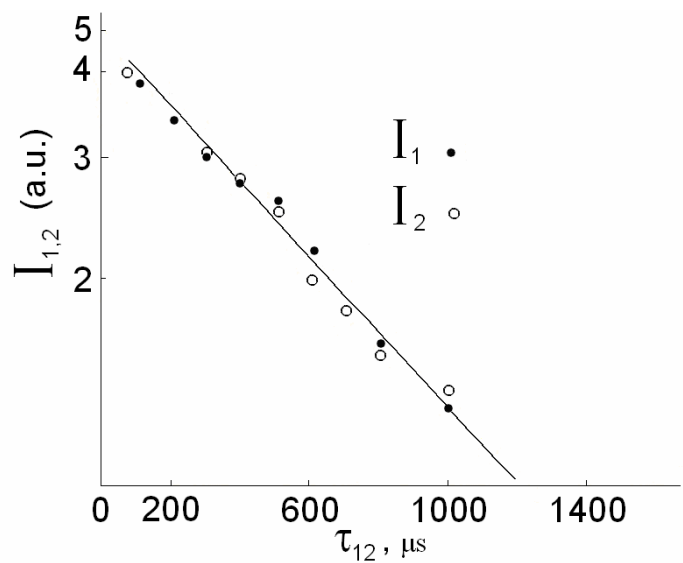

Fig.9.

FIGURE CAPTIONS

Fig.1. Oscillograms of the SPE and its secondary signals from $^7$Fe nuclei in lithium ferrite. The lower beam indicates the position and duration of the RF pulses (T=77 K, $f_{NMR}$=71, 6 MHz, $\tau$=10 $\mu$s , T=300 $\mu$s ).

Fig.2. Oscillograms of the TPE and its secondary signals in lithium ferrite. The lower beam shows the position and duration of the RF pulses (T=77 K, $f_{NMR}$=71, 6 MHz, $\tau_1$=1.8 $\mu$s , $\tau_2$=2.2 $\mu$s , $\tau_{12}$=5 $\mu$s , T=300 $\mu$s ).

Fig.3. The intensity of (1) SPE signal and (2) its secondary signal as functions of the repetition period of RF pulses T in lithium ferrite.

Fig.4. The intensity of (1) TPE and (2), (3) its secondary signals as functions of the period of repetition of the RF pulses T in lithium ferrite.

Fig.5. SPE signal (1-3) in lithium ferrite at the influence of RF pulse repetition period showing the process of complete disappearance of the SPE signal when T becomes more than $\sim$ 1 sec (3).

Fig.6. The intensity of (1) SPE and (2) TPE as functions of the period of repetition of the RF pulses T in the intermetallic compound $Co_2MnSi$. (T=77 K, $f_{NMR}$=71, 6 MHz, $\tau$=10 $\mu$s , T=300 $\mu$s ).

Fig.7. The intensity of (1) SPE and (2) TPE as functions of the repetition period of the RF pulses in hexagonal cobalt.

Fig.8. Intensities of SPE main and secondary signals as functions of the duration of RF pulse $\tau$ in lithium ferrite at $f_{NMR}$=74 MHz

Fig.9. Intensities of TPE main and secondary signals as functions of the interval between RF pulses $\tau_{12}$ in lithium ferrite at $f_{NMR}$=74 MHz.